\documentclass[12pt]{article}
\usepackage{amssymb}
\usepackage{hyperref}
\usepackage[pdftex]{graphicx}

\newcommand{\la}{\lambda}
\newcommand{\om}{\omega}
\newcommand{\ep}{\varepsilon}
\newcommand{\dl}{\delta}
\newcommand{\Tr}{{\rm Tr}}
\newcommand{\rme}{{\rm e}}
\newcommand{\rmi}{{\rm i}}

\newtheorem{theorem}{Theorem}

\begin{document}
\title{Quantum control landscape for a $\Lambda$-atom in the vicinity of second
order traps}
\author{Alexander Pechen$^{1,2}$\thanks{\href{mailto:pechen@mi.ras.ru}{pechen@mi.ras.ru}; \href{http://www.mathnet.ru/eng/person/17991}{www.mathnet.ru/eng/person/17991}}\,\, and David J. Tannor$^1$\thanks{\href{mailto:david.tannor@weizmann.ac.il}{david.tannor@weizmann.ac.il}; \href{http://www.weizmann.ac.il/chemphys/tannor}{www.weizmann.ac.il/chemphys/tannor}}}
\date{}
\maketitle

\vspace{-1cm}
\begin{center}
$^1$Weizmann Institute of Science, Rehovot 76100, Israel\\
$^2$Steklov Mathematical Institute of Russian Academy of Sciences,\\ Gubkina str. 8, 119991, Moscow, Russia
\end{center}

\begin{abstract} We show that the second order traps in the control landscape
for a three-level $\Lambda$-system found in our previous work {\it Phys. Rev.
Lett.} {\bf 106}, 120402 (2011) are not local maxima: there exist directions in
the space of controls in which the objective grows. The growth of the objective
is slow --- at best 4th order for weak variations of the control. This implies
that simple gradient methods would be problematic in the vicinity of second
order traps, where more sophisticated algorithms that exploit the higher order
derivative information are necessary to climb up the control landscape
efficiently. The theory is supported by a numerical investigation of the
landscape in the vicinity of the $\ep(t)=0$ second order trap, performed using
the GRAPE and BFGS algorithms.
\end{abstract}

\section{Introduction}
Manipulation of atomic systems via specially tailored laser pulses is an
important application of quantum control
theory~\cite{Tannor1985,Rice2000,Butkovskiy1990,Brumer2003,Tannor2007,
Letokhov2007,Sklarz2004,Brif2010,Poulsen2010}. One of the questions of current
interest is the nature of the control landscape that describes the value of the
objective as a functional of the control field $\ep(t)$, in particular the
nature of the critical points on the landscape. For the objective $J(\ep)$ a
critical control field $\ep$ is a {\it trap} of the landscape if $\ep$ is a
local but not a global maximum. Control landscapes for closed quantum systems
were studied in~\cite{Landscapes1,Landscapes2}, where the absence of traps was
shown under some assumptions. The analysis for open quantum systems was
performed in~\cite{Pechen2008,Wu2008} and later was extended to a unified
analysis of control landscapes for open classical and quantum
systems~\cite{Pechen2010}.

In this work we consider critical points for control objectives of the form
\begin{equation}\label{eq3} J(\ep)=\Tr
[U^{\vphantom{\dagger}\ep}_T\rho_0U^{\dagger\ep}_T O] \end{equation} for a
3-level $\Lambda$-system isolated from the environment. Such objectives describe
a wide variety of quantum control phenomena, e.g., breaking a desired chemical
bond, producing selective atomic and molecular excitations, etc. Here
$\rho_0=|i\rangle\langle i|$ is the initial system state which is assumed to be
pure, $O$ is the target observable, and $U^\ep_T$ is the evolution operator
describing the evolution of the system from the initial time $t=0$ to the final
time $T>0$ under the action of coherent control field $\ep(t)$. The evolution
operator satisfies the equation
\[
 \frac{dU^\ep_t}{dt}=-i(H_0+V\ep(t))U^\ep_t,\qquad U^\ep_0=\mathbb I
\]
where $H_0=\sum_{j=1}^3\ep_j |j\rangle\langle j|$ is the free system Hamiltonian
and $V$ is the operator describing coupling of the system to the control field.

A control field $\ep$ is a {\it second order trap} for the objective functional
$J(\ep)$ if the gradient $\nabla J_\ep=0$ and the Hessian $H_\ep=\dl^2
J/\dl\ep^2$ is negative semi-definite but $J(\ep)$ is not a global maximum. More
generally, a control field $\ep$ is an {\it $n$-th order trap} (here $n$ is an
even natural number) if it is not a global maximum and $\dl
J_\ep:=J(\ep+\dl\ep)-J(\ep)= R(\dl\ep)+O(\dl\ep^{n+1})$, where the functional
$R\ne 0$, $R(\dl\ep)=O(\dl\ep^n)$, and $R(\dl\ep)\le 0$.

Various algorithms have been used to find optimal controls in quantum systems
including gradient search, e.g. GRAPE~\cite{GRAPE} and its modifications for
rapidly time-varying Hamiltonians~\cite{Motzoi2011}, Krotov-type
methods~\cite{Krotov1983,Tannor1992,Krotov2009,Zhu1998,Maday2003}, the
Broyden-Fletcher-Goldfarb-Shanno (BGFS) algorithm and its
modifications~\cite{B,F,G,S}, genetic algorithms and evolutionary
strategies~\cite{Judson2002,Pechen2006}, and combined
approaches~\cite{Eitan2011,Machnes2011}. The analysis of the existence or
absence of traps, including $n$-order traps, is important for determining proper
algorithms for finding optimal control fields. In the absence of traps, local
(e.g., gradient) search should generally be able to find globally optimal
controls (exceptions may occur if the initial control for a gradient-based
search is chosen exactly at a saddle point, where the gradient of the objective
is zero). If the landscape has second order or $n$-order traps then, since the
gradient of the objective at a critical point is zero and in addition, at
second-order or $n$-order traps the Hessian is negative semi-definite, the
objective in their vicinity may grow not faster than at a third order in the
small variation $\dl\ep$ of the control. Therefore the search for globally
optimal controls with simple gradient methods would be problematic in a vicinity
of second-order or $n$-order traps, where more sophisticated algorithms
exploiting the higher order information about the objective are necessary to
efficiently climb up the control landscape.

Second order traps were shown to exist for a general class of quantum
systems~\cite{PechenTannor2011}. The simplest example is a three-level
$\Lambda$-system, where zero control field was shown to be a second order trap.
In this work we continue the analysis of~\cite{PechenTannor2011} and show that
this second order trap is not a local maximum; there exist a direction in which
the objective grows. We also perform a numerical study of the landscape in a
vicinity of the $\ep(t)=0$ second order trap.

\section{Three-level $\Lambda$-system}
\begin{figure}
\begin{center}
\includegraphics[scale=0.7]{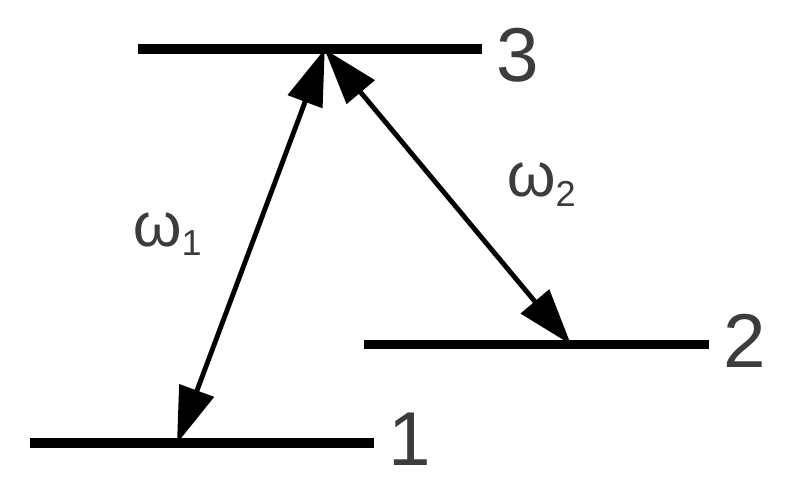}
\caption{\label{fig1}The simplest example of a quantum system possessing a
second order trap is a three-level $\Lambda$-system initially in the ground
state. The control field $\ep(t)=0$ is a second order trap for maximizing
expectation of any target operator of the form
$O=\sum_{i=1}^3\la_i|i\rangle\langle i|$ with $\la_2>\la_1>\la_3$. However, this
field is not a trap, as shown in Theorem~\ref{T2}.}
\end{center}
\end{figure}

The simplest example where second order traps appear is the following problem:
maximizing the expectation of an operator $O=\sum_{j=1}^3\la_j|j\rangle\langle
j|$ with $\la_2>\la_1>\la_3$ for a three-level non-degenerate $\Lambda$-atom
which is initially in the ground state $\rho_0=|1\rangle\langle 1|$
(Fig.~\ref{fig1}). The interaction Hamiltonian $V$ for $\Lambda$-atom satisfies
$V_{12}=0$, which is consistent with the controllability assumption if
$V_{13}\ne 0$ and $V_{23}\ne 0$. Note that the pure state controllability for
$N$-level systems with odd number of levels requires producing the full
unitary group $su(N)$, while for $N$ even the symplectic group $sp(N/2)$ alone
is
sufficient~\cite{Brockett72,Alessandro2003}. The two
allowed transition frequencies are
$\om_1=\ep_3-\ep_1$ and $\om_2=\ep_3-\ep_2$; non-degeneracy implies that
$\om_1\ne \om_2$. Globally optimal control fields are those which steer
$|1\rangle$ completely into $|2\rangle$ producing the global maximum of the
objective with the objective value $J_{\rm max}=\la_2$.

Without loss of generality we can set $\la_1=0$ and $\la_2=1$ (replacing $O$ by
$O'=( O-\la_1\mathbb I)/(\la_2-\la_1)$ and noticing that linear transformations
of the target observable do not affect the properties of critical points). Since
$\la_3<\la_1=0$, we can set $\la_3=-\la$, where $\la>0$. After these
transformations the objective $J$ takes the form
\begin{equation}\label{eq1:0}
 J(\ep)=P_{1\to 2}(\ep)-\la P_{1\to 3}(\ep),\qquad \la>0
\end{equation}
where $P_{i\to f}(\ep)=|\langle f|U_T^\ep|i\rangle|^2$ is the transition
probability from the state $|i\rangle$ to the state $|f\rangle$ ($i=1$,
$f=2,3$). The objective takes the values in the interval $-\la\le J\le 1$.

If $\ep(t)=0$, then the system remains in the ground state and therefore both
transition probabilities are zero, $P_{1\to 2}(0)=P_{1\to 3}(0)=0$. The
corresponding objective value $J=0$ is neither a global maximum nor a global
minimum.

\begin{theorem}
The control field $\ep(t)=0$ is a second order trap with the objective value
$J=0<J_{\rm max}$.
\end{theorem}

More generally, a control $\ep(t)=\ep_0$ is a second order trap if the initial
density matrix and target operator have the form $\rho_0=|\tilde k\rangle\langle
\tilde k|$ and $O= \sum_{i=1}^n \la_i|\tilde i\rangle\langle \tilde i|$ in the
basis $|\tilde i\rangle$ of the modified Hamiltonian $\tilde H_0=H_0-\mu\ep_0$,
where $1<k<n$ and $\la_1>\la_2>\dots>\la_n$, and the dipole moment satisfies
$\langle \tilde i|\mu|\tilde k\rangle=0$ for all $i<k$.
This important result was proved in~\cite{PechenTannor2011}. However, previously
it was not known if these second order traps are true traps (local maxima). The
following theorem states that it is not the case and there exist local
directions in which the objective increases. The analysis of the general case is
equivalent to the case $\ep(t)=0$ with the modified Hamiltonian $\tilde H_0$,
therefore only the case $\ep(t)=0$ will be treated in the rest of the
manuscript.

\begin{theorem}\label{T2}
The control field $\ep(t)=0$ is not a local maximum; there exists a direction
$\dl\ep$ in which the objective $J(\dl\ep)$ grows as $J\approx
\|\dl\ep\|^4$.
\end{theorem}
\noindent {\bf Proof.} Matrix elements determining the transition probabilities
under the action of a small $\dl\ep$ can be expanded in powers of $\dl\ep$
as
\begin{eqnarray}
 \langle 3| U^{\dl\ep}_T|1\rangle &=& A_1+A_2+A_3+O(\dl\ep^4)\label{eq1:1}\\
 \langle 2| U^{\dl\ep}_T|1\rangle &=& B_2+O(\dl\ep^3)\label{eq1:2}
\end{eqnarray}
where the terms $A_n$ and $B_n$ are of the order $O(\dl\ep^n)$. Their explicit
form can be derived from the perturbation expansion
\[
 U^{\dl\ep}_T={\rm e}^{-iTH_0}\left(\mathbb I+\sum\limits_{n=1}^\infty
(-i)^n\int_0^T
dt_1\int_0^{t_1} dt_2\dots \int_0^{t_{n-1}} dt_n
\dl\ep(t_1)\dots\dl\ep(t_n) V_{t_1}\dots V_{t_n}\right)
\]
for the evolution operator $U^{\dl\ep}_T$ (where $V_t=e^{it H_0}Ve^{-it H_0}$)
as:
\begin{eqnarray}
A_1&=&
 -i\rme^{-iT\ep_3}\int_0^T dt_1\dl\ep(t_1)\langle3|V_{t_1}|1\rangle=
 -iV_{31}\rme^{-iT\ep_3}\int_0^T dt_1\dl\ep(t_1){\rm e}^{it\om_1}\label{eq2:1}\\
A_2&=&
 -\rme^{-iT\ep_3}\int_0^T dt_1\int_0^{t_1}dt_2\dl\ep(t_1)\dl\ep(t_2)
 \langle 3|V_{t_1}V_{t_2}|1\rangle\label{eq2:2}\\
A_3&=&
 i\rme^{-iT\ep_3}\int_0^T dt_1\int_0^{t_1}dt_2\int_0^{t_2}dt_3\dl\ep(t_1)
 \dl\ep(t_2)\dl\ep(t_3)\langle 3|V_{t_1}V_{t_2}V_{t_3}|1\rangle\label{eq2:3}\\
B_2&=&
 -\rme^{-iT\ep_2}\int_0^T dt_1\int_0^{t_1}dt_2\dl\ep(t_1)\dl\ep(t_2)
 \langle 2|V_{t_1}V_{t_2}|1\rangle\label{eq2:4}
\end{eqnarray}

Note that $\langle 3|V_{t_1}V_{t_2}|1\rangle=0$ and thus $A_2\equiv 0$ for any
$\dl\ep$. Substituting eqs.~(\ref{eq1:1}) and~(\ref{eq1:2}) into
eq.~(\ref{eq1:0}) produces the
following expansion for the objective around zero control field:
\begin{eqnarray*}
 \dl J_{\ep=0}&=&J(\dl\ep)-J(0)\\
 &=&-\la |A_1|^2+\Bigl(|B_2|^2-2\la{\rm Re}(A_1^*A_3)\Bigr)+O(\dl\ep^6)
\end{eqnarray*}
(If the diagonal elements of $V$ are zero as is normally the case for the
$\Lambda$-system, this implies that for the $\Lambda$-system $\langle
2|V_{t_1}V_{t_2}V_{t_3}|1\rangle=0$ and
$\langle 3|V_{t_1}V_{t_2}V_{t_3}V_{t_4}|1\rangle=0$. Therefore the term of
fifth order on the right hand side of this equation vanishes and the
remainder is of sixth order.)

If $A_1\ne 0$, then $\dl J=-\la |A_1|^2+O(\dl\ep^4)<0$ for sufficiently small
$\dl\ep$. The Hessian of the objective at $\ep=0$ is negative
semi-definite and therefore $\ep=0$ is a second order trap. However, if there
would exist a $\dl\ep$ such that $A_1=0$ and $B_2\ne 0$, then would be $\dl J=
|B_2|^2+O(\dl\ep^6)>0$ and $\ep=0$ would not be a trap.

An example of $\dl\ep$ for which $A_1=0$ and $B_2\ne 0$ is the following (we
assume $T\ge T' :=2\pi/\om_1$). Set $\dl\ep(t)=\alpha\chi_{[0,T']}(t)$, where
$\chi_{[0,T]}$ is the characteristic function of the segment $[0,T]$, i.e.
$\dl\ep(t)=\alpha$ if $0\le t\le T'$ and
$\dl\ep(t)=0$ if $t>T'$ ($\alpha>0$ is a small number). Then
\begin{eqnarray*}
 A_1&=&\alpha\frac{V_{31}}{\om_1}\Bigl(1-\rme^{\rmi
       T'\om_1}\Bigr)\rme^{-iT\ep_3}=0\\
 B_2&=&-V_{23}V_{31}\int_0^{T'} dt_1\rme^{\rmi t_1\om_1}
       \dl\ep(t_1)\int_0^{t_1} dt_2\rme^{-\rmi t_2\om_2}\dl\ep(t_2)\\
    && =\alpha^2 V_{23}V_{31}\frac{\rme^{2\pi\rmi r}-1}{\om_2(\om_2-\om_1)},
       \qquad r=\frac{\om_2}{\om_1}
\end{eqnarray*}
The quantity $B_2\ne 0$ if $r\notin\mathbb Z$ which is trivially satisfied
since $0<\om_2<\om_1$. This proves the Theorem.

The most general conditions for $\dl\ep$ to increase the objective at the 4th
order are
\begin{eqnarray}
&& \tilde\dl\ep_1(\om_1):=\int_0^T dt\rme^{it\om_1} \dl\ep(t)=0\\
&& \tilde\dl\ep_2(\om_1,\om_2):=\int_0^T dt_1\rme^{\rmi
t_1\om_1}\dl\ep(t_1)\int_0^{t_1}dt_2\rme^{-\rmi t_2\om_2}\dl\ep(t_2)\ne 0
\end{eqnarray}
The condition $\tilde\dl\ep(\om_1)=0$ means that a control field $\dl\ep$ which
increases the objective should have zero amplitude at the frequency $\om_1$ of
the $1\to 3$ transition. Such a field does not populate the level $|3\rangle$ in
the first order of the perturbation theory (via one-photon processes). Because
of the special form of the interaction Hamiltonian $V$, the level $|3\rangle$
can not be populated also in the second order of the perturbation theory (via
two-photon processes). However, level $|2\rangle$ can be populated at
second order in the perturbation even by fields which have zero amplitude
at the frequency $\om_1$ if $\tilde\dl\ep_2(\om_1,\om_2)\ne 0$. Such a transfer
of population increases the objective at the rate $\sim\dl\ep^4$.
Similar control pulses which are non-resonant but have non-zero two-photon
transition probability were used for coherent control of multiphoton
transitions~\cite{Meshulach1998}.

The analysis shows that for the $\Lambda$-system in some directions around
second order traps the objective increases. Such behavior was also found for a
four-level system by constructing a second order trap where the
objective increases in a direction
$\delta\ep(t)=\chi_{[0,\pi]}(t)$~\cite{Schirmer2010} (Example~2).
However, this increase is slow (at best it is of 4th order in $\dl\ep$ in our
case, and of 3rd order in the example of~\cite{Schirmer2010}) and simple
gradient methods may not be effective in escaping these critical points.
Therefore more sophisticated algorithms exploiting the higher order information
about the objective are generally necessary to climb up the landscape in the
vicinity of second order traps.

\section{Numerical analysis}
In this section we consider piecewise constant controls of the form
\[
 \ep(t)=\sum\limits_{k=1}^M c_k\chi_{[t_k,t_{k+1}]}(t)
\]
where $\chi_{[t_k,t_{k+1}]}(t)$ is the characteristic function of the interval
$[t_k,t_{k+1}]$ (i.e., $\chi_{[t_k,t_{k+1}]}(t)=1$ if $t\in[t_k,t_{k+1}]$ and
$\chi_{[t_k,t_{k+1}]}(t)=0$ otherwise), $t_k=\Delta t(k-1)$ and $\Delta t =T/M$.
In this
representation the control is an $M$-dimensional vector $C=(c_1,\dots, c_M)$.
The objective is defined as
\[
 J(C)=\Tr[ U_T\rho_0 U^\dagger_T O],\qquad \textrm{ where } U_T=U_T(C)
\]
The gradient of the objective is
\[
 \nabla J(C)=\left (\frac{\partial J(C)}{\partial c_1},\dots, \frac{\partial
J(C)}{\partial c_M} \right)
\]
Here the partial derivative with respect to $c_l$ has the form
\[
\frac{\partial J(C)}{\partial c_l}=2\Delta t\cdot {\rm Im}\Bigl(
\Tr\Bigl[W^\dagger_l V W_l \rho_0 U^\dagger_T O U_T\Bigr]\Bigr)
\]
where
\[
W_l = U_l U_{l-1}\dots U_2 U_1,\qquad
U_k=\rme^{-\rmi (H_0+c_k V)\Delta t}, \qquad U_T=W_M
\]

We use GRAPE~\cite{GRAPE} and BFGS from the
MATLAB Optimization Toolbox as local search algorithms. The GRAPE algorithm
starts with a randomly generated
initial control $C_1$, computes the gradient $\nabla J(C_1)$, updates the
control as $C_2=C_1+\epsilon\nabla J(C_1)$, where $\epsilon>0$ is a small
number, and continues the loop either until the objective reaches the value
$J_{\rm stop}=1-I_{\rm err}$, where $I_{\rm err}$ is some threshold or until
some number $K_{\rm stop}$ of iterations is reached. The components of
the initial control $C_1$ are generated randomly with uniform distribution in
the interval $[-c_0, c_0]$ with some $c_0>0$. Since the second order trap is at
$\ep(t)=0$, $c_0$ can be viewed as measure of distance from the second order
trap.

\begin{figure}
\begin{center}
\includegraphics[scale=.7]{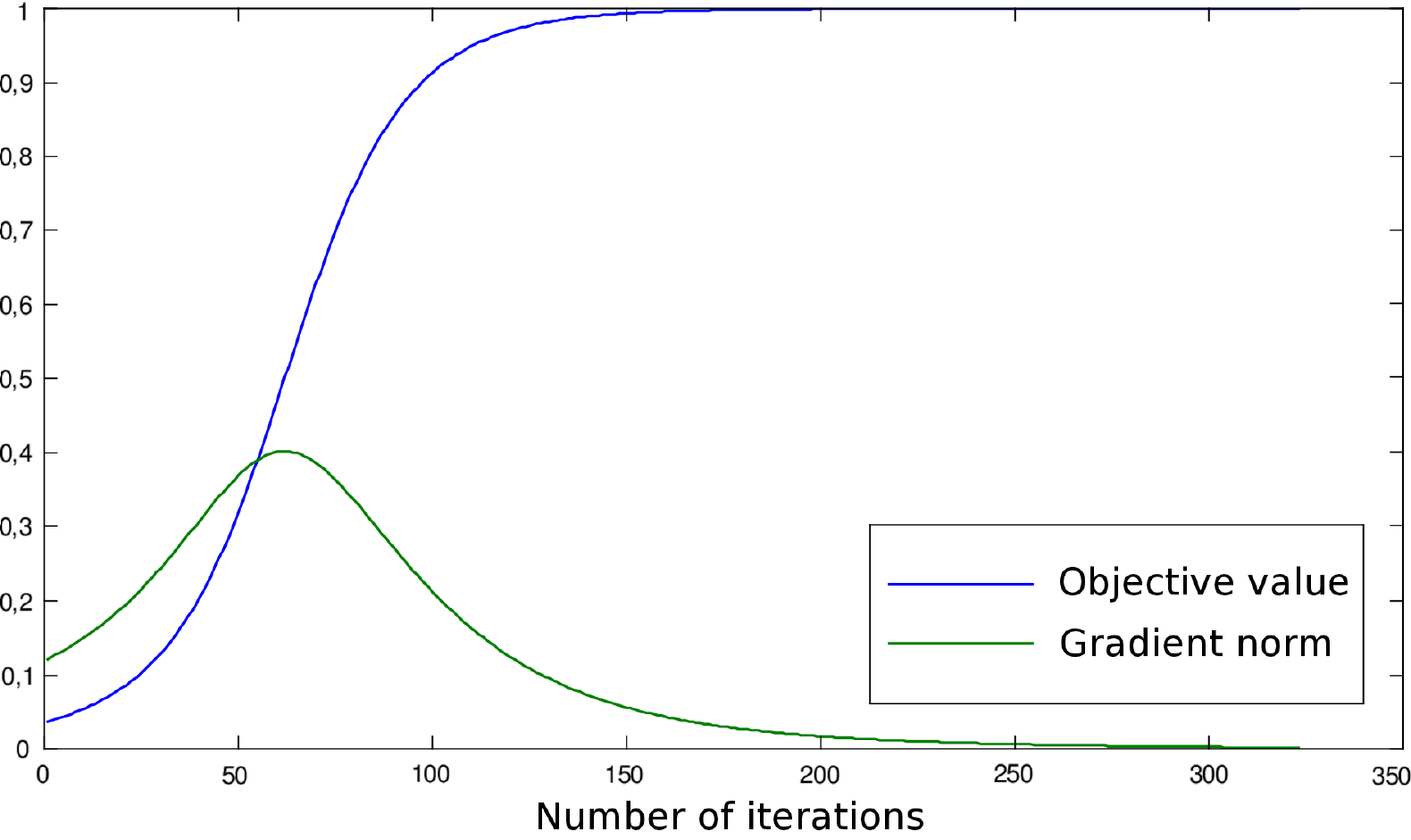}
\caption{\label{fig2} (Color online) A typical example of the behavior of the
objective and the norm of the gradient vs number of iterations for maximizing
the $P_{1\to 2}$ transition probability for a three-level $\Lambda$-system. The
parameters are: $M=200$, $c_0=1$, $\epsilon=0.1$, $I_{\rm err}=10^{-5}$.
The objective value $0.99999$ is reached in $323$ iterations.}
\end{center}
\end{figure}

\begin{figure}
\begin{center}
\includegraphics[scale=.9]{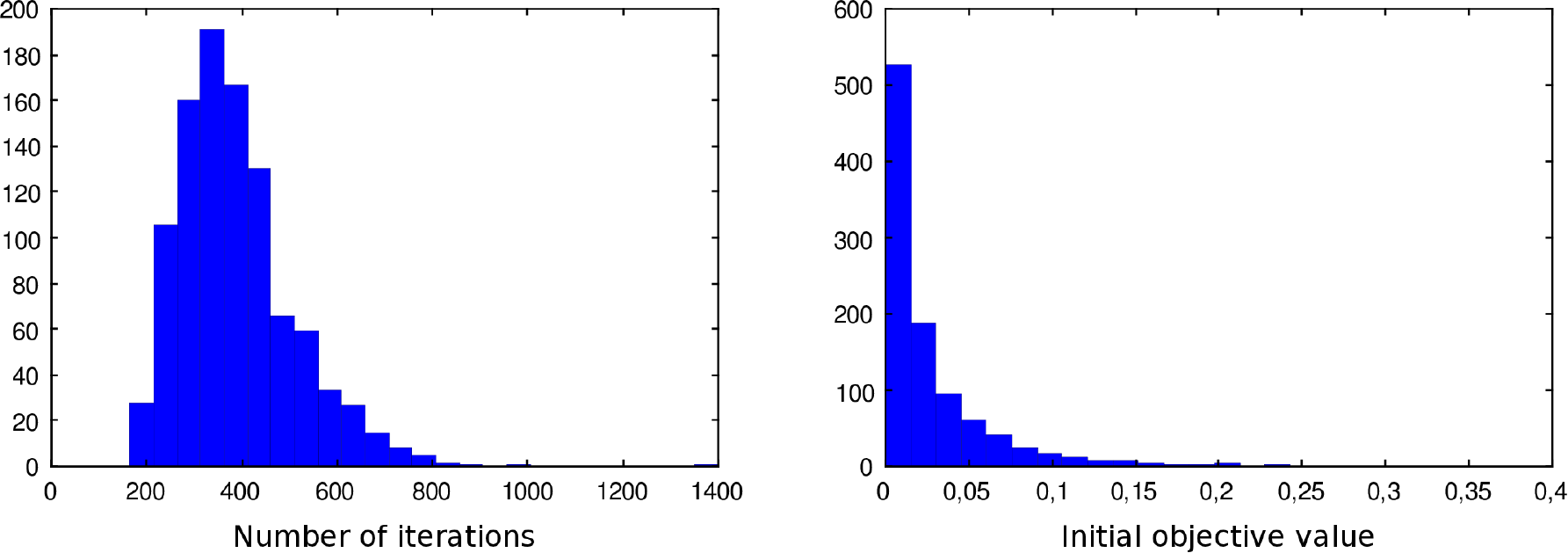}
\caption{\label{fig3} (Color online) $L=1000$ initial controls were randomly
generated to produce these plots for maximizing $P_{1\to 2}$ in a three-level
$\Lambda$-system ($M=200$, $c_0=1$, $\epsilon=0.1$, $I_{\rm err}=10^{-5}$). Left
histogram represents the distribution of the number of iterations required to
reach the objective value $J=1-I_{\rm err}=0.99999$ (minimum number of
iterations
is $165$, maximum $1400$, mean $388$, and the standard deviation
$\sigma=125$). Right histogram represents the distribution of the initial
objective values. In each histogram $25$ intervals were used.}
\end{center}
\end{figure}

We consider the following parameters of the $\Lambda$-system:
\[
H_0=\left(\begin{array}{ccc}0&0&0\\0&1&0\\0&0&2.5\end{array}\right), \qquad
V=\left(\begin{array}{ccc}0&0&1\\0&0&1.7\\1&1.7&0\end{array}\right)
\]
The final time in the simulations is chosen as $T=10$, that is several times
larger than typical oscillation periods in the system. First we consider the
case $\la=0$ which corresponds to the standard $P_{1\to 2}$ problem, where the
objective is to transfer the population from the ground state $|1\rangle$ to the
intermediate state $|2\rangle$. For this case, second order traps were not
discovered in~\cite{PechenTannor2011}. Figure~\ref{fig2} shows that for this
case it is easy to find, even in a small number of iterations, control fields
that produce an objective which deviates from the global maximum by as little as
$10^{-5}$. Figure~\ref{fig3} (left) shows the distribution of the number of
iterations required to achieve $99.999\%$ population transfer and (right) the
distribution of initial objective values. It is seen that despite the fact that
the search generally starts from low objective values, i.e. the bottom of the
landscape, it always reaches the top in a small number of iterations.

Now we consider the case $\la>0$, with the goal to analyze how for small initial
controls the $\ep(t)=0$ second order trap may significantly slow down the search
for global maxima. Recall that $O=|2\rangle\langle 2|-\la|3\rangle\langle 3|$,
and therefore larger values of $\la$ should have a more significant effect on
the slowdown of the search. We choose $\la=5$ and perform the optimization for
$c_0=0.1, 0.2, 0.3,\dots 1$. For each choice of $c_0$, $L=100$ runs of the GRAPE
algorithm were performed with components of the initial controls generated
randomly in the interval $[-c_0, c_0]$. Other parameters were chosen as follows:
$K_{\rm stop}=1000$ for the maximum allowed number of iterations, $I_{\rm
err}=0.1$ for the allowed deviation from the global maximum of the objective,
$\epsilon=0.1$ for the iteration size, $T=10$ for the final time, and $M=200$
for the number of components in each control.

\begin{figure}
\begin{center}
\includegraphics[scale=.8]{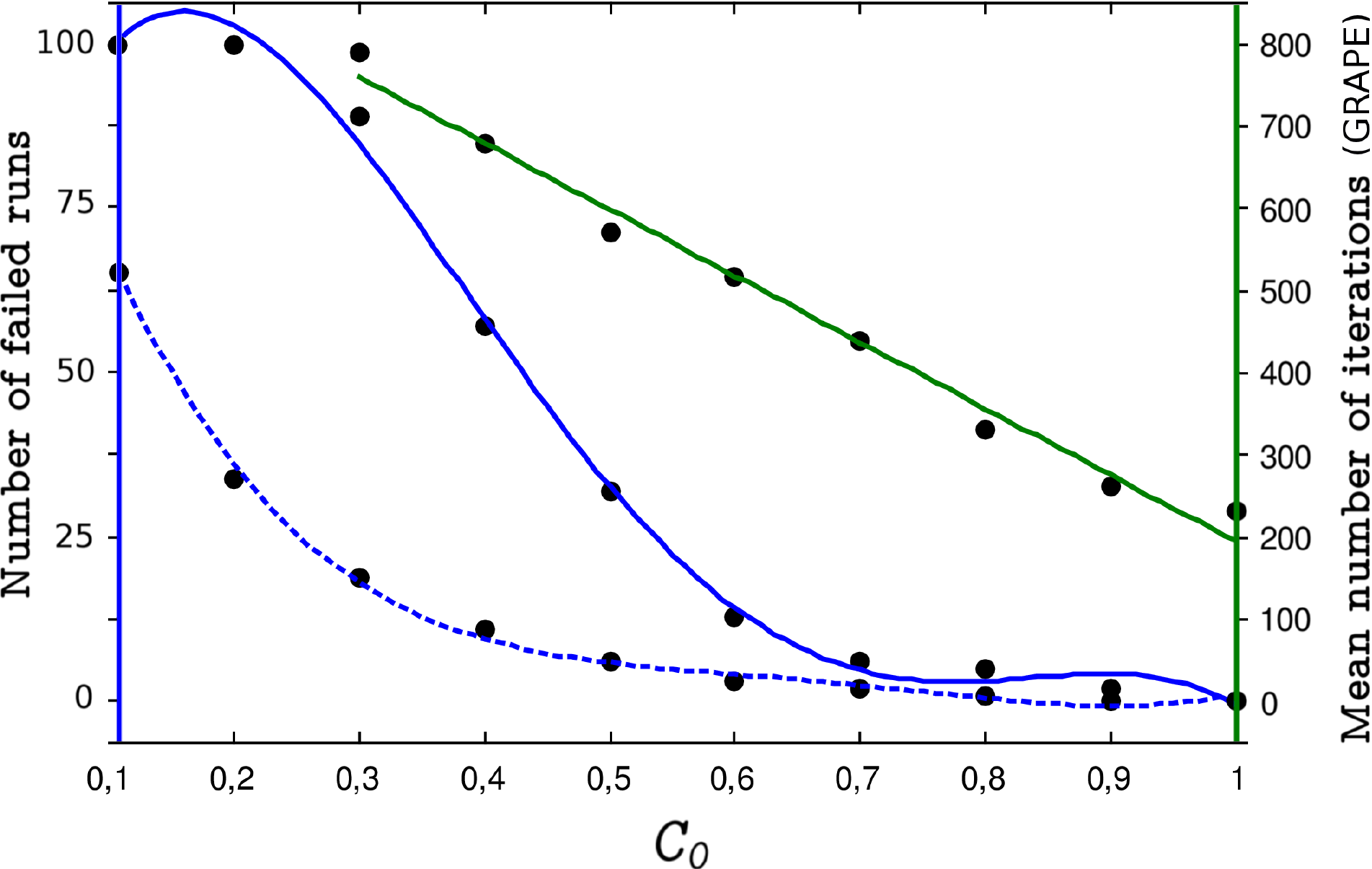}
\caption{\label{fig4} (Color online) For each point, $L=100$ runs of either
GRAPE (with $\epsilon=0.1$) or BFGS were performed with initial controls
generated randomly in the interval $[-c_0, c_0]$ ($T=10$, $M=200$, and $I_{\rm
err}=0.1$). The bottom two (blue) lines show the dependence of the number of
failed runs for GRAPE (solid line, using $K_{\rm stop}=1000$) and BFGS (dashed
line, with the termination condition determined by the default MATLAB criterion)
plotted with a fourth order polynomial interpolations. Upper (green) line shows
the mean number of iterations for those runs of GRAPE that were successful, i.e.
that found controls that achieve the fidelity $0.9$ in fewer than $K_{\rm
stop}=1000$ iterations (linear interpolation).}
\end{center}
\end{figure}

For each $c_0$ we determine the number of unsuccessful attempts $N_{\rm fail}$
(among $L$ runs) when the algorithm fails to reach the objective value $1-I_{\rm
err}$ in less than $K_{\rm stop}$ iterations. We also determine the mean number
of iterations required to reach the objective value $1-I_{\rm err}$ in the
successful $L-N_{\rm fail}$ runs. These data are plotted in Fig.~\ref{fig4}. For
$c_0=0.1$ and $c_0=0.2$ we find that $N_{\rm fail}=L$, showing that {\it all}
runs of the algorithm fail if the amplitudes $c_i$ of the initial control
satisfy $c_i\le 0.2$ ($i=1,\dots, M$). The number of failed runs decreases with
increase of $c_0$ and becomes close to zero when $c_0=1$. Starting from
$c_0=0.3$, the mean number of iterations required to reach the desired objective
value decreases linearly. We also observe that if the limiting number of
iterations $K_{\rm stop}$ is increased then some runs will be successful even
for small $c_0$. However, in practice there always exist limits on the number of
possible iterations and therefore if a large number of iterations is required to
reach a desired objective value, for practical purposes this may look like a
trap. Another comment is that the number of iterations might be dependent on the
details of the algorithm. For example, using an adaptive step size may improve
performance of the gradient search at points with small gradient. For
comparison we use the BFGS method. The number of fails for every
$c_0$ is less than for GRAPE, particularly for small $c_0$, as plotted by the
dashed line in Fig.~\ref{fig4}.
The decrease in the efficiency of the linear gradient search was also observed
for quantum systems which have forbidden transitions between distant energy
levels~\cite{Moore2011}. The three-level $\Lambda$-system has a forbidden
$1\to2$ transition and therefore is within this class. Our work however studies
a different dependence of the efficiency --- on $c_0$, which is effectively the
distance of the initial control from the second order trap.

\section*{Conclusions} This work shows that constant controls, which are
second order traps in the control landscape for a three-level $\Lambda$-system,
are not local maxima: there exists directions around these controls in which the
objective increases at 4th order. A numerical investigation shows that indeed
the search with simple gradient methods becomes slow in a sufficiently small
neighborhood of $\ep(t)=0$. These results suggest that in the vicinity of second
order traps, simple gradient algorithms may not be sufficient, and more
sophisticated algorithms that exploit higher order derivative information may
be necessary.

\section*{Acknowledgments} A. Pechen acknowledges support of the Marie Curie
International Incoming Fellowship within the 7th European Community Framework
Programme. This research was supported by the Minerva Foundation and is made
possible in part by the historic generosity of the Harold Perlman family.


\begin{thebibliography}{99}
\bibitem{Tannor1985} D.J.~Tannor, S.A.~Rice, J.~Chem.~Phys. {\bf 83}, 5013
(1985).
\bibitem{Rice2000} S.A.~Rice, M.~Zhao, {\it Optical Control of
Molecular Dynamics}, Wiley, New York, 2000.
\bibitem{Butkovskiy1990} A.G. Butkovskiy, Yu.I.~Samoilenko, {\it Control of
Quantum-Mechanical Processes and Systems}, Kluwer Academic, Dordrecht, 1990.
\bibitem{Brumer2003} P.W. Brumer, M. Shapiro, {\it Principles of the Quantum
Control of Molecular Processes}, Wiley-Interscience, 2003.
\bibitem{Tannor2007} D.J. Tannor, {\it Introduction to Quantum Mechanics: A
Time Dependent Perspective}, University Science Press, Sausalito, 2007.
\bibitem{Letokhov2007} V.S. Letokhov, {\it Laser Control of Atoms and
Molecules}, Oxford University Press, New York, 2007.
\bibitem{Sklarz2004} S.E. Sklarz, D.J. Tannor, N. Khaneja, Phys. Rev. A
{\bf 69}, 053408 (2004).
\bibitem{Brif2010} C. Brif, R. Chakrabarti, H. Rabitz, New J. Phys. {\bf
12}, 075008 (2010).
\bibitem{Poulsen2010} U.V. Poulsen, S. Sklarz, D.J. Tannor, T. Calarco,
Phys. Rev. A {\bf 82}, 012339 (2010).
\bibitem{Landscapes1} H. Rabitz, M. Hsieh, C. Rosenthal, Phys. Rev. A {\bf
72}, 052337 (2005).
\bibitem{Landscapes2} M. Hsieh, R. Wu, C. Rosenthal, H. Rabitz, J. Phys. B:
At. Mol. Opt. Phys. {\bf 41}, 074020 (2008).
\bibitem{Pechen2008} A. Pechen, D. Prokhorenko, R. Wu, H. Rabitz, J. Phys.
A: Math. Theor., {\bf 41}, 045205 (2008).
\bibitem{Wu2008} R. Wu, A. Pechen, H. Rabitz, M. Hsieh, B. Tsou, J. Math.
Phys. {\bf 49}, 022108 (2008).
\bibitem{Pechen2010} A. Pechen, H. Rabitz, Europhysics Letters {\bf 91},
60005 (2010).
\bibitem{GRAPE} N. Khaneja, T. Reiss, C. Kehlet, T. Schulte-Herbr\" uggen, S.
Glaser, J. Magn. Reson. {\bf 172}, 296 (2005).
\bibitem{Motzoi2011} F. Motzoi, J.M. Gambetta, S.T. Merkel, F.K.
Wilhelm, Phys. Rev. A {\bf 84}, 022307 (2011).
\bibitem{Krotov1983} V.F. Krotov, I. N. Fel'dman, Eng. Cybernetics {\bf 17},
123 (1983).
\bibitem{Tannor1992} D.J. Tannor, V. Kazakov, V. Orlov, in {\it Time-Dependent
Quantum Mechanics}, edited by J. Broeckhove and L. Lathouwers, Plenum Press,
New York, 1992, pp. 347--360.
\bibitem{Krotov2009} V.F. Krotov, Automation and Remote Control, {\bf 70(3)},
357 (2009).
\bibitem{Zhu1998} W. Zhu and H. Rabitz, J. Chem. Phys. {\bf 109}, 385 (1998).
\bibitem{Maday2003} Y. Maday and G. Turinici, J. Chem. Phys. {\bf 118}, 8191
(2003)
\bibitem{B} C.G. Broyden, J. Inst. Math. Appl. {\bf 6}, 76 (1970).
\bibitem{F} R.A. Fletcher {\it Comput. J.} {\bf 13}, 317 (1970).
\bibitem{G} D.A. Goldfarb, {\it Math. Comput.} {\bf 24}, 23 (1970).
\bibitem{S} D.F. Shanno, {\it Math. Comput.} {\bf 24}, 647 (1970).
\bibitem{Judson2002} R.S. Judson and H. Rabitz, Phys. Rev. Lett. {\bf 68}, 1500
(1992).
\bibitem{Pechen2006} A. Pechen, H. Rabitz, Phys. Rev. A {\bf 73}, 062102
(2006).
\bibitem{Eitan2011} R. Eitan, M. Mundt, D.J. Tannor, Phys. Rev. A {\bf 83},
053426 (2011).
\bibitem{Machnes2011} S. Machnes {\it et al}, Phys. Rev. A {\bf 84}, 022305
(2011).
\bibitem{PechenTannor2011} A.N. Pechen, D.J. Tannor, Phys. Rev. Lett.
{\bf 106}, 120402 (2011).
\bibitem{Brockett72} R. Brockett, SIAM J. Applied Math. {\bf vol. 25(2)},
213--225 (1973).
\bibitem{Alessandro2003} F. Albertini, D. D'Alessandro, IEEE Trans. Autom.
Control {\bf 48}, 1399 (2003).
\bibitem{Meshulach1998} D. Meshulach, Y. Silberberg, Phys. Rev. A {\bf 60},
1287 (1999).
\bibitem{Moore2011} K.W. Moore, R. Chakrabarti, G. Riviello, H. Rabitz,
Phys. Rev. A {\bf 83}, 012326 (2011).
\bibitem{Schirmer2010} P. de Fouquieres, S.G.~Schirmer, {\it Quantum Control
Landscapes: A Closer Look}, Preprint arXiv:1004.3492 (2010).
\end{thebibliography}
\end{document}